\documentclass[11pt,a4paper]{article}
\usepackage{jheppub_kim}

\usepackage{pdflscape}
\usepackage{amsmath}
\usepackage{amssymb}
\usepackage{dcolumn}
\usepackage{bm}
\usepackage{color}
\usepackage{epsfig}
\usepackage{amsfonts}
\usepackage{graphicx}
\usepackage{subfigure}
\usepackage{dcolumn}

\begin{document}

\title{P-V criticality of  first-order entropy corrected  AdS black holes in massive gravity}
\author[a]{ S. Upadhyay,}
\author[b]{B. Pourhassan  }
\author[b]{and H. Farahani}

\affiliation[a]{Centre for Theoretical Studies,
Indian Institute of Technology Kharagpur,  Kharagpur-721302, WB, India}
 \affiliation[b]{School of Physics, Damghan University, Damghan, Iran}

\emailAdd{sudhakerupadhyay@gmail.com}
 \emailAdd{b.pourhassan@du.ac.ir}
 \emailAdd{h.farahani@umz.ac.ir}

\abstract{We consider a massive black hole in four dimensional AdS space  and study the effect of thermal fluctuations on the thermodynamics of the black hole. We consider thermal fluctuations as logarithmic correction  terms in the entropy. We analyse the effect of logarithmic correction on thermodynamics potentials like Helmholtz and Gibbs which are found decreasing functions. We study critical points and stability and find that presence of logarithmic correction is necessary to have stable phase and critical point.}

 \keywords{Thermodynamics, Critical behavior, Massive  black hole.}

\maketitle

\section{Introduction}
In order to prevent the violation of the second law of thermodynamics one can associate a maximum entropy with black holes \cite{1, 1a, 2, 4, 4a}. Otherwise, when an object with a finite entropy crossed the horizon, the entropy of the universe would spontaneously reduce  and, therefore, violets the second law of thermodynamics.
The scaling of the mentioned maximum entropy  with the black hole horizon area led to the holographic principle \cite{5, 5a}, which equates the degrees of freedom in any region of space to the degrees of freedom of the boundary. The holographic principle will be corrected near the Planck scale, as well as quantum gravity corrections modify the topology of 
space-time at this scale \cite{6, 6a}. As we know, the holographic principle is inspired by the entropy-area relation, hence the quantum gravity corrections will modify the entropy-area relation. In that case, the original black hole entropy is given by  $S_0= A/4$, where $A$ is the black hole event horizon area. Then, the corrected entropy-area relation of a black hole may be written as
$S=  S_0 + \alpha \log A+  \gamma_1 A^{-1} + \gamma_2 A^{-2} \cdots$, where $\alpha, \gamma_1, \gamma_2 \cdots$,
are the coefficients which depend on the black hole parameters. Also, the area dependence has been obtained by the specific models of the quantum gravity. In that case, the logarithmic correction of the form $\alpha \log A$  has  already been used to study the corrected thermodynamics of some kinds of black holes such as G\"{o}del like black hole \cite{godel}. We should note that the thermodynamics corrections of black holes can be studied by using the non-perturbative quantum general
relativity \cite{1z}, where the conformal blocks of the conformal field theory  are used to study the behavior of the
density states. The effects of quantum corrections to the black hole thermodynamics have already been studied with help of the Cardy formula \cite{card}.  The corrected  thermodynamics of a black hole are also studied under the effect of matter fields around a black hole \cite{other, other0, other1}. The thermodynamics corrections  produced by the string theory have also been studied which are in agreement with the other approaches to quantum gravity \cite{solo1, solo2, solo4, solo5}.
The  corrections to the thermodynamics of a  dilatonic black hole have also been discussed 
and  observed to have the same universal manner \cite{jy}. The partition
function of a black hole is very useful to study the corrected thermodynamics
of a black hole \cite{bss}. It is also possible to use the generalized uncertainty principle to produce thermodynamics corrections,
which yields to the logarithmic correction \cite{mi, r1}, in agreement with the other approaches to quantum
gravity. It should be noted that the  Einstein equations in the Jacobson formalism are thermodynamics identities \cite{z12j, jz12}. Therefore, a quantum correction to the space-time topology would produce thermal fluctuations in the black holes thermodynamics, and it has the same universal form as expected from the quantum gravitational effects \cite{l1, SPR, more}.

In fact, these corrections have already been  considered to study several black geometries.
For example, an AdS charged black hole has been studied under the effect of logarithmic correction of the entropy 
and  has been found that the thermodynamics of the AdS black hole is modified
due to the thermal fluctuations \cite{1503.07418}. The effect of thermal fluctuations on the
thermodynamics a black Saturn have also been studied in the Ref. \cite{1505.02373}. It has
been found that the thermal fluctuations do not have any major effect on the stability of the black Saturn. The  thermal fluctuations for a
 modified Hayward black hole have been studied, where it  has been found that thermal fluctuations reduce the pressure
and internal energy of the Hayward black hole \cite{1603.01457}. The effect of thermal fluctuations on the thermodynamics of a
 charged dilatonic black saturn has also  been studied
 \cite{1605.00924}. It was stated that the thermal fluctuations can be studied
 either using a conformal field theory or using the fluctuations in the energy
 of this system. However, it has been found that the  fluctuations in the energy and the conformal field theory produce the same results  for a charged dilatonic black saturn. This result may differ for the other black objects. Thermodynamics of a small singly spinning Kerr-AdS black hole under the effects of thermal fluctuations has been studied recently  \cite{NPB} with the conclusion that the logarithmic correction becomes important when the size of the black hole is sufficient small, which enable us to test the effects of quantum fluctuations
on the black holes by analyzing the effects of thermal fluctuations for example on
dumb holes (the analogous for black holes) to obtain the correct coefficient for
the correction terms \cite{Annals}. Such corrections may affect the critical behaviors of black object, for example a dyonic charged anti-de Sitter black hole, which is holographic dual of a Van der Waals fluid \cite{Hen1}, is  considered in the Ref. \cite{PRD} where logarithm-corrected thermodynamics is investigated with the result that holographic picture is still valid. However,  the van der Waals phase
transitions of charged black holes in massive gravity without any order correction is discussed in Ref. \cite{Hen2}.
 An important application of such logarithmic correction can be found as the study of  quark-gluon plasma properties by using AdS/CFT correspondence \cite{JHEP,CAN,EPJC2,G}. It may, 
for example, affect the shear viscosity to entropy ratio \cite{EPJC}. 

Massive gravity, overcoming its traditional problems, has found a resurgence of interest due to recent progress \cite{curt}, yielding an avenue for addressing  the
cosmological constant naturalness problem. The possibility of a massive graviton has been studied  first by Fierz and Pauli \cite{fir,fir1}. Further, van Dam and Veltman
\cite{van} and Zakharov \cite{zak} had found the linear theory
coupled to a source  which discuss the
curious fact that the theory makes predictions different from
those of linear gravity theory even in the limit as the graviton mass goes to
zero. Later, 
some specific nonlinear massive gravity theories have been studied \cite{bd,bd1} which possess a ghostlike instability, known as the
Boulware-Deser ghost. The significant progress has been made in
  construction of the massive gravity theories without such instability \cite{dr,dr1}.
  The most straightforward way to
construct the massive gravity theories is to simply add a mass term to the
GR action, giving the graviton a mass in such a way that GR is recovered as mass vanishes.
Recently, a charged BTZ black holes in the context of massive gravity's rainbow 
has been studied \cite{sud}. The massive BTZ black holes in the presence of Maxwell and 
Born-Infeld  electrodynamics in asymptotically (A)dS spacetimes is studied \cite{Hen}.
 More recently,  the higher order correction of the entropy and 
 the  thermodynamical properties of   Schwarzschild-Beltrami-de Sitter black 
hole are studied \cite{sud1}. In fact, the $P-V$ criticality of charged black holes in 
 Gauss-Bonnet-massive gravity is also presented \cite{hen1}. The   van der Waals like phase transition \cite{bibh} and 
$P-V$ criticality of AdS black holes in a general framework \cite{bibh1} are recently discussed.

Now, we would like to obtain the effect of the first-order (logarithmic) corrected entropy on the thermodynamics and $P-V$ criticality of black holes in AdS space-time of massive gravity. For this purpose, we  consider  the 4-dimensional  charged black
hole  in massive gravity
with a negative cosmological constant and discuss  the effect of 
 first-order correction on various thermodynamics quantities. For example, we derive the  entropy, 
Hawking temperature, Helmholtz function,  internal energy, pressure, enthalpy and Gibbs free energy. We analyse the
Helmholtz free energy   with respect to correction coefficient 
$\alpha$, which confirms that the effect of the logarithmic correction
is important at small $r_+$ (or high temperature) and there exists a critical radius
for which Helmholtz free energy  vanishes.
Also, we show that the logarithmic correction has no important effect on the pressure
of the black hole with large event horizon radius.
In fact,  the  internal energy, enthalpy and Gibbs free energy are found a decreasing function of correction parameter. We further discuss the holographic duality of logarithmic corrected AdS black hole in massive gravity with  
Van der Waals fluid  for the large black hole and find that
the thermal fluctuations have no important effect. In order to study the
effect of thermal fluctuations on the critical points, we analyse $P-V$ behavior of
the black hole. We  discuss the effect of thermal fluctuations in view of critical point  and  stability of the model.
From the plot, we  find that the logarithmic correction
will be helpful to remove instability of the black hole. For the stability of the model, we obtain a necessary requirement that the trace of Hessian matrix of the Helmholtz free energy must be non-negative.

This paper is organized as follow. In the next section, we recall the black holes of AdS space-time in massive gravity. In section 3, we introduce logarithmic corrected entropy as leading order of thermal fluctuations. In section 4, we discuss about holographic dual picture of the black hole. In section 5, we study critical point and stability of the black hole. Finally, in section 6, we discuss conclusion and summarize the results.

\section{AdS black holes in massive gravity}
Let us consider the following action for (3+1)-dimensional
massive gravity with a Maxwell field  \cite{cai}
\begin{equation}\label{1}
S=\frac{1}{k^2}\int d^4x \sqrt{-g}\left[R-2\Lambda -\frac{1}{4}F^2+m^2\sum_i^4 c_i{\cal U}_i\right],
\end{equation}
where $\Lambda=-3/l^2$ is the cosmological constant and $k = 1, 0$, or $−1$, correspond to a
sphere, Ricci flat, or hyperbolic horizon for the black hole,
respectively. Here  $F_{\mu\nu}$ is the Maxwell field-strength tensor, $c_i$ are constants, and ${\cal U}_i$ are symmetric
polynomials of the eigenvalues of the  matrix $\sqrt{g^{\mu\alpha}f_{\alpha\nu}}$, where $f_{\mu\nu}$ is   a fixed
symmetric tensor. 

The action admits a static black hole solution with the
space-time metric and reference metric as
\begin{equation}\label{2}
ds^2=-f(r)dt^2+\frac{dr^2}{f(r)}+r^2h_{ij}dx^idx^j,\ \ i,j=1,2.
\end{equation}
Here, $h_{ij}dx^idx^j$ is the   line
element for an Einstein space with constant curvature. The metric function $f(r)$ in terms of electric charge $q$  is written by \cite{cai}
\begin{equation}\label{3}
f(r)=k+\frac{r^2}{l^2}-\frac{m_0}{r}+\frac{q^2}{4r^2}+\frac{m^2  c_1 r}{2}
+ m^2 c_2.
\end{equation}
 The black hole horizon can be determined by  setting $f(r)|_{r=r_+}=0$,
hence the mass parameter $m_0$ which is related to the total mass of the black hole is given by
\begin{equation}\label{4}
m_{0}=(k+m^2 c_2)r_{+}+\frac{r_{+}^3}{l^2}+\frac{q^2}{4r_{+}}+\frac{m^2  c_1 r_{+}^{2}}{2},
\end{equation}
where outer horizon $r_{+}$ is largest real root of the equation $f(r)=0$, ie,
\begin{equation}\label{5}
k+\frac{r^2}{l^2}-\frac{m_0}{r}+\frac{q^2}{4r^2}+\frac{m^2  c_1 r}{2}
+ m^2 c_2=0.
\end{equation}
For example by choosing the parameter as $m_{0} = 2$, $c_{1} = 1$, $c_{2} = 1$, $k = 1$, $l = 1$, $m = 0.2$, and $q = 1$ one can obtain $r_{-}=0.1346110283$ while $r_{+}=0.9126757206$ together two complex roots. There is also a chemical potential corresponding to the electrical charge $q$ given by,
\begin{equation}\label{6}
\mu=\frac{q}{r_{+}}.
\end{equation}

\section{Fisrt-order corrected thermodynamics}
\begin{figure}[h!]
 \begin{center}$
 \begin{array}{cccc}
\includegraphics[width=50 mm]{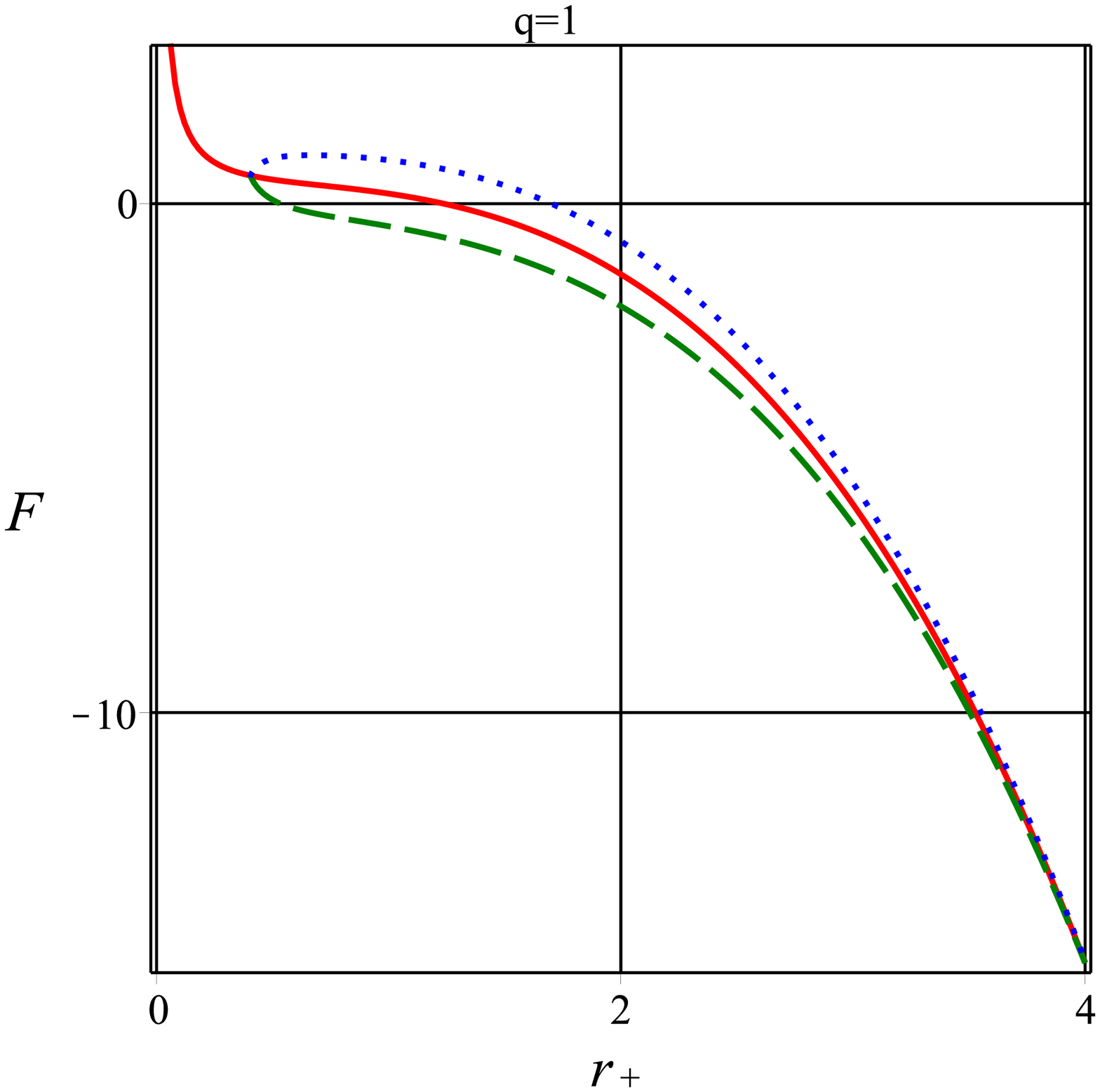}\includegraphics[width=50 mm]{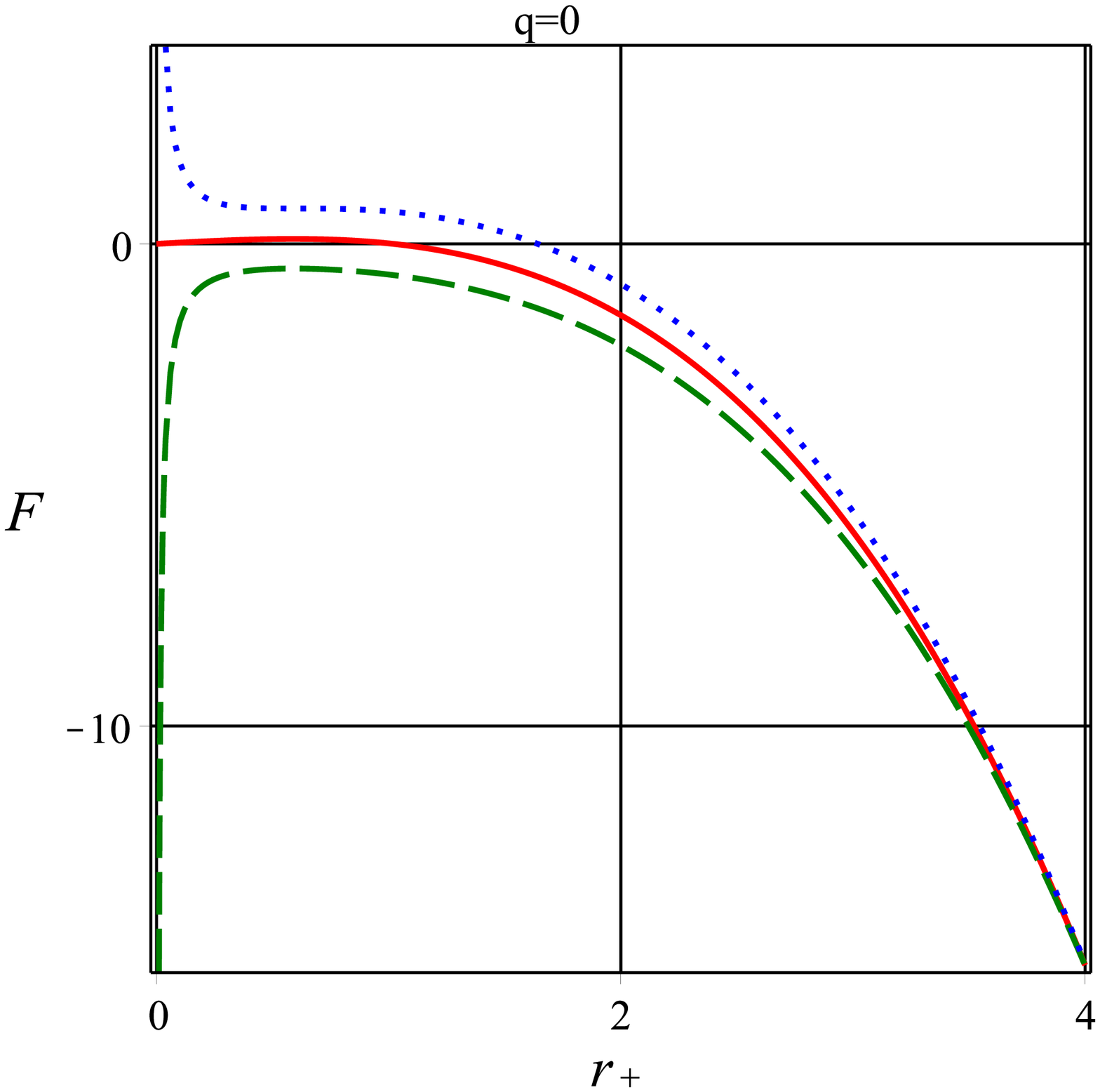}
 \end{array}$
 \end{center}
\caption{Helmholtz free energy in terms of $r_{+}$ for $m=0.2$ and we set
unit values for all  other parameters. $\alpha=0,1,-1$ are denoted by solid red, dash green, and dotted blue, respectively. Left plot with $q=1$ and right plot with $q=0$.}
 \label{fig1}
\end{figure}

The first order corrected entropy is given by \cite{l1},
\begin{equation}\label{s}
S=S_0-\frac{\alpha}{2}\log(S_0T_H^2),
\end{equation}
where $\alpha$ is a constant having dimension of length
and the zeroth order entropy $S_0$ is given by \cite{xu},
 \begin{eqnarray}\label{s0}
 S_0= {\pi} r_+^2.
 \end{eqnarray}
Using the definition of Hawking temperature with relation to the surface gravity on
the outer horizon $r_+$ \cite{cai},
 \begin{eqnarray}\label{t}
T_H=\frac{1}{4\pi}\left[\frac{k}{r_+}+\frac{3r_+}{l^2}-\frac{1}{4}\frac{q^2}{r_+^3} +m^2 c_1+\frac{m^2c_2}{r_+}\right].
 \end{eqnarray}
 Exploiting relations (\ref{s}) and (\ref{t}), the  corrected entropy is given by,
 \begin{eqnarray}\label{s1}
S= {\pi}r_+^{2}  - {\alpha} \log \left[\frac{1}{4\sqrt{\pi}}\left(k+3\frac{r^2_+}{l^2}-
\frac{q^2}{4r^2_+}+m^2c_1r_++m^2c_2\right)\right],
\end{eqnarray}
Using the entropy and temperature, we can find the Helmholtz function,
\begin{eqnarray}\label{10}
F=-\int SdT_H,
\end{eqnarray}
as follow
\begin{eqnarray}\label{ss}
F=k\frac{r_+}{4} -\frac{r_+^3}{4l^2}+\frac{3q^2}{16r_+}+\frac{1}{4}m^2c_2r_{+}+F_{\alpha},
\end{eqnarray}
where
\begin{eqnarray}\label{12}
F_{\alpha}&=&\frac{\alpha}{24\pi l^2}\left(\frac{l^2q^2}{r_+^3}-36r_+\right)\nonumber\\
&+&\frac{\alpha}{16\pi l^2}\left(12r_{+}-\frac{q^2l^2}{r_+^3}+\frac{4kl^2}{r_+}+\frac{4c_2l^2m^2}{r_+} \right)\nonumber\\ &\times&\log\left[\frac{1}{4\sqrt{\pi}}\left(k+c_2m^2-\frac{q^2}{4r_+^2}+m^2c_1r_+ +\frac{3r_+^2}{l^2}\right) \right]\nonumber\\
&+&\frac{\alpha c_1m^2}{4\pi}\log\left[\frac{12r_+}{l_p}-\frac{q^2l^2}{r_+^3l_p}+4\frac{l^2k}{r_+l_p}+4\frac{l^2m^2(c_2+c_1r_+)}{r_+l_p} \right],
\end{eqnarray}
and $l_p$ is some constant of integration and has dimension as length. In the Fig. \ref{fig1} we draw Helmholtz free energy in terms of horizon radius with variation of correction coefficient $\alpha$. As expected, $F_{\alpha}\rightarrow0$ as $r_{+}\gg1$, which means that the effect of the logarithmic correction is important at small $r_{+}$. In the right plot of the Fig. \ref{fig1}, we can see the uncharged ($q=0$) case, for which  the effect of logarithmic correction becomes significant at high temperature (infinitesimal $r_{+}$). It should be noted that the cases of $k=0$ and $k=\pm1$ lead to the similar result. Also,
the  negative values of $c_{1}$ and $c_{2}$ have no important effect. Left plot of the Fig. \ref{fig1} shows that there is a critical radius $r_{c}$ where $F_{\alpha}=F_{-\alpha}=0$ and we have $r_{-}\leq r_{c} \leq r_{+}$, where equality holds for the extremal black hole ($m_{0}\approx q$ with $r_{+}\approx 0.4$). It should be noted that the value of the event horizon  depends on the value of $m_{0}$.\\

Also, the second term of the rhs of Eq. (\ref{ss}) corresponds
to an ordinary AdS black hole with thermodynamics
pressure related to the cosmological constant,
\begin{eqnarray}\label{13}
P_\Lambda(q=m=\alpha=0)=-\frac{\Lambda}{16\pi}= \frac{3}{16\pi}\frac{1}{l^2},
\end{eqnarray}
and thermodynamic volume,
\begin{eqnarray}\label{14}
V=\frac{4}{3}\pi r_+^3.
\end{eqnarray}

In order to calculate the internal energy, we use the well-known
thermodynamics relation $E=F+TS$, and obtain
\begin{equation}\label{15}
E= k\frac{r_+}{2}+\frac{r_+^3}{2l^2}+\frac{ q^2}{8r_+}+\frac{1}{2}m^2c_2r_++
\frac{1}{4}m^2c_1r_+^2 +E_{\alpha}
\end{equation}
where
\begin{eqnarray}\label{16}
E_{\alpha}&=&\frac{\alpha c_1m^2}{4\pi}\log\left[\frac{16\sqrt{\pi}l^2}{l_p}\left( \frac{ 12r_+^4-q^2l^2+4kl^2r_+^2
 +4m^2l^2r_+^2(c_2+c_1r_+) }{ 4kl^2r_+^3+4c_2m^2l^2r_+^3-q^2l^2r_++4m^2c_1l^2r_+^3+12r_+^5 } \right)\right]\nonumber\\
 &+&\frac{\alpha}{24\pi l^2}\left(\frac{l^2q^2}{r_+^3}-36r_+\right).
\end{eqnarray}
It is clear that the correction parameter $\alpha$ decreases the value of the internal energy for all cases of $k=0,\pm1$, i.e., the charged, uncharged and extremal black hole, respectively.\\
As we know, modified pressure due to
thermal fluctuation can be obtained using the derivative of
the Helmholtz function with respect to the volume,
\begin{equation}\label{17}
P=-\left(\frac{\partial F}{\partial V}\right)_T
\end{equation}
which gives
\begin{equation}\label{18}
P=-\frac{k}{16\pi r_+^2}+\frac{3}{16\pi l^2} +\frac{3q^2}{64\pi r_+^4}-\frac{m^2c_2}{16\pi r_+^2}+P_{\alpha},
\end{equation}
where
\begin{eqnarray}\label{19}
P_{\alpha}&=&\frac{3\alpha}{8\pi^2 r_+^2 l^2}+\frac{\alpha q^2}{32\pi^2r_+^6}-\frac{\alpha}{64\pi^2 r_+^2l^2}\left(12 \frac{3q^2l^2}{r_+^4} -\frac{4kl^2}{r_+^2}
-\frac{4c_2l^2m^2}{r_+^2} \right) \nonumber\\
&\times&\log\left[\frac{1}{4\sqrt{\pi}}\left(k+c_2m^2-\frac{q^2}{4r_+^2}+m^2c_1r_+ +\frac{3r_+^2}{l^2}\right) \right]\nonumber\\
&-&\frac{\alpha}{32\pi^2 r_+^2l^2}\left[\frac{(l^2q^2-4r_+^2(kl^2+c_2l^2m^2+3r_+^2))(8r_+^4+q^2l^2+2c_1m^2l^2r_+^3)}{-8r_+^8+l^2r_+^4(q^2-4kr_+^2-4m^2c_2r_+^2-4m^2c_1r^3_+)} \right]   \nonumber\\
&- & \frac{\alpha c_1m^2}{16\pi^2r_+^2} \left[\frac{12r_+^4+l^2(3q^2-4kr_+^2-4c_2m^2r_+^2)}{12r_+^5-q^2l^2r_++4l^2r_+^3(k+m^2c_2+m^2c_1r_+)}\right].
\end{eqnarray}
As expected, pressure is decreasing function of $r_{+}$ while it is increasing function of $\alpha$ for small radius. We find that logarithmic correction has no important effect on the pressure of the black hole with large event horizon radius.\\
Also, we can obtain enthalpy as,
\begin{eqnarray}
H=E+PV =\frac{3}{4}\frac{r_+^3}{l^2}+\frac{5}{12}kr_++\frac{3}{16}\frac{q^2}{r_+}+
\frac{5}{12}m^2c_2r_++\frac{1}{4}m^2c_1r_+^2+  H_\alpha,
\end{eqnarray}
where
\begin{eqnarray}
H_\alpha &=&   \frac{\alpha q^2}{12\pi r_+^3}-\frac{\alpha r_+}{\pi l^2}- \frac{\alpha c_1m^2r_+}{12\pi} \left[\frac{12r_+^4+l^2(3q^2-4kr_+^2-4c_2m^2r_+^2)}{12r_+^5-q^2l^2r_++4l^2r_+^3(k+m^2c_2+m^2c_1r_+)}\right]\nonumber\\
 &+&\frac{\alpha c_1m^2}{4\pi}\log\left[\frac{16\sqrt{\pi}l^2}{l_p}\left( \frac{ 12r_+^4-q^2l^2+4kl^2r_+^2
 +4m^2l^2r_+^2(c_2+c_1r_+) }{ 4kl^2r_+^3+4c_2m^2l^2r_+^3-q^2l^2r_++4m^2c_1l^2r_+^3+12r_+^5 } \right)\right] \nonumber\\
&-&\frac{\alpha r_+}{24\pi  l^2}\left[\frac{(l^2q^2-4r_+^2(kl^2+c_2l^2m^2+3r_+^2))(8r_+^4+q^2l^2+2c_1m^2l^2r_+^3)}{-8r_+^8+l^2r_+^4(q^2-4kr_+^2-4m^2c_2r_+^2-4m^2c_1r^3_+)} \right]   \nonumber\\
&-&\frac{\alpha r_+}{48\pi  l^2}\left(12 +\frac{3q^2l^2}{r_+^4} -\frac{4kl^2}{r_+^2}
-\frac{4c_2l^2m^2}{r_+^2} \right)\log\left[\frac{1}{4\sqrt{\pi}}\left(k+c_2m^2-\frac{q^2}{4r_+^2}\right.\right.\nonumber\\
&+& \left.\left. m^2c_1r_+ + \frac{3r_+^2}{l^2}\right) \right].
\end{eqnarray}
We find that enthalpy in decreasing function of $\alpha$ as well.\\
Gibbs free energy using the
relation $G=H-TS=F+PV$ is obtained as,
\begin{eqnarray}
G=\frac{1}{6}kr_++\frac{q^2}{4r_+}+\frac{1}{6}m^2c_2r_++G_\alpha,
\end{eqnarray}
where
\begin{eqnarray}
G_\alpha &=&   \frac{\alpha q^2}{12\pi r_+^3}-\frac{\alpha r_+}{\pi l^2}- \frac{\alpha c_1m^2r_+}{12\pi} \left[\frac{12r_+^4+l^2(3q^2-4kr_+^2-4c_2m^2r_+^2)}{12r_+^5-q^2l^2r_++4l^2r_+^3(k+m^2c_2+m^2c_1r_+)}\right]\nonumber\\
&+&\frac{\alpha c_1m^2}{4\pi}\log\left[\frac{12r_+}{l_p}-\frac{q^2l^2}{r_+^3l_p}+4\frac{l^2k}{r_+l_p}+4\frac{l^2m^2(c_2+c_1r_+)}{r_+l_p} \right]+ \frac{\alpha r_+}{24\pi l^2}\left(12
\right.\nonumber\\
&-&\left.\frac{3q^2l^2}{r_+^4}   +\frac{8kl^2}{r_+^2}+\frac{8c_2l^2m^2}{r_+^2} \right) \log\left[\frac{1}{4\sqrt{\pi}}\left(k+c_2m^2-\frac{q^2}{4r_+^2}+m^2c_1r_+ +\frac{3r_+^2}{l^2}\right) \right]\nonumber\\
&-&\frac{\alpha r_+}{24\pi  l^2}\left[\frac{(l^2q^2-4r_+^2(kl^2+c_2l^2m^2+3r_+^2))(8r_
+^4+q^2l^2+2c_1m^2l^2r_+^3)}{-8r_+^8+l^2r_+^4(q^2-4kr_+^2-4m^2c_2r_+^2-4m^2c_1r^3_+)}
\right],
\end{eqnarray}
which is decreasing function of $\alpha$ like other thermodynamics potentials.
 \section{Holographic duality}
\begin{figure}[h!]
 \begin{center}$
 \begin{array}{cccc}
\includegraphics[width=50 mm]{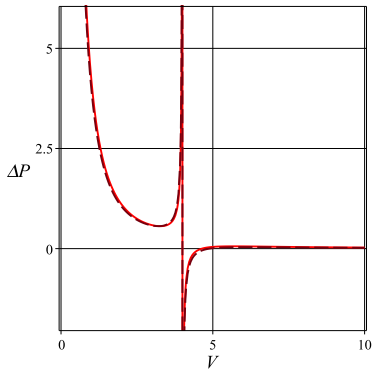}
 \end{array}$
 \end{center}
\caption{ $\Delta P=P-P_{W}$ in terms of $V$ for $m=0.2$, $a=b=4$ and $l=2$, and we set
unit values for all other parameters.  Blue dashed line corresponding to $q=1$ while solid red line corresponding to $q=0$.}
 \label{fig2}
\end{figure}

It will be interesting if   AdS black hole in massive gravity has a holographic dual of the form of Van der Waals fluid with the following equation of state
\begin{equation}
(P_{W}+\frac{a}{V^2})(V-b)=T,
\end{equation}
where we assumed $K_{B}=1$ (unit of Boltzmann constant). Also, $a$ and $b$ are some positive constants in which the constant $a$ parameterizes the strength of the intermolecular interactions, while the constant $b$ accounts for the volume excluded owing to the finite size of molecules in fluid. If $a$ and $b$ are both set to zero, the equation of state for an ideal gas can be recovered. It means that
\begin{equation}
P_{W}=\frac{T}{V-b}-\frac{a}{V^2}.
\end{equation}
Now,   AdS black hole in massive gravity with logarithmic correction is holographic dual of Van der Waals fluid if $P=P_{W}$, where $P$ given by the equation (\ref{18}). By using numerical analysis we find that the mentioned duality holds for the large black hole ($V\gg1$) and therefore thermal fluctuations have no important effect in this case. In the Fig. \ref{fig2}, we draw $\Delta P=P-P_{W}$ and find some regions where $\Delta P=0$ corresponding to the large $V$. In this limit, the value of $\alpha$ is not important and thermal fluctuations have no key role to violate holographic dual picture. There exists a divergency  also at critical volume. Hence, it is possible to have dual Van der Waals fluid in presence of logarithmic correction. In order to find the effect of thermal fluctuations on the critical points, we should analyze $P-V$ behavior of the black hole.

\begin{figure}[h!]
 \begin{center}$
 \begin{array}{cccc}
\includegraphics[width=50 mm]{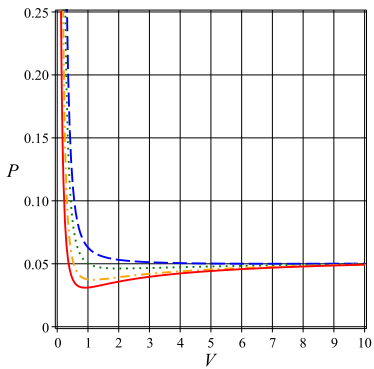}
 \end{array}$
 \end{center}
\caption{Pressure in terms of $V$ for $m=0.2$, $q=0.5$, and we set
unit values for all other parameters.  Blue dashed line corresponding to $\alpha=1$, green dotted line corresponding to $\alpha=0.6$, orange dash dotted line corresponding to $\alpha=0.2$, solid red line corresponding to $\alpha=0$.}
 \label{fig3}
\end{figure}

Therefore, we can study $P-V$ criticality via the following relations:
\begin{eqnarray}\label{cr}
 \left(\frac{\partial P}{\partial V} \right)_{T=T_c}&=&0,\nonumber\\
 \left(\frac{\partial^2 P}{\partial V^2} \right)_{T=T_c}&=&0.
\end{eqnarray}
By using the relations (\ref{14}) and (\ref{18}), we can draw $P$ in terms of $V$ as illustrated by the Fig. \ref{fig3}. It is clear that critical point exists also in presence of logarithmic correction. Typical behavior of $P$ for the selected values of parameter shows critical point at $V\approx3.5$ which means $T_{c}\approx0.3$. At the $\alpha=0$ limit, one can obtain the following condition to have the first condition of (\ref{cr}):
\begin{equation}\label{cr1}
V_{c}=\frac{\pi q^{3} \sqrt{6}}{(c_{2}m^{2}+k)^{\frac{3}{2}}},
\end{equation}
while the second condition gives,
\begin{equation}\label{cr2}
V_{c}=\frac{7\sqrt{35}}{25}\frac{\pi q^{3} \sqrt{6}}{(c_{2}m^{2}+k)^{\frac{3}{2}}}.
\end{equation}
It is clear that both equations (\ref{cr1}) and (\ref{cr2}) never satisfy simultaneously. It means that without thermal fluctuations there is no critical point.

\section{Critical points and stability}
\begin{figure}[h!]
 \begin{center}$
 \begin{array}{cccc}
\includegraphics[width=40 mm]{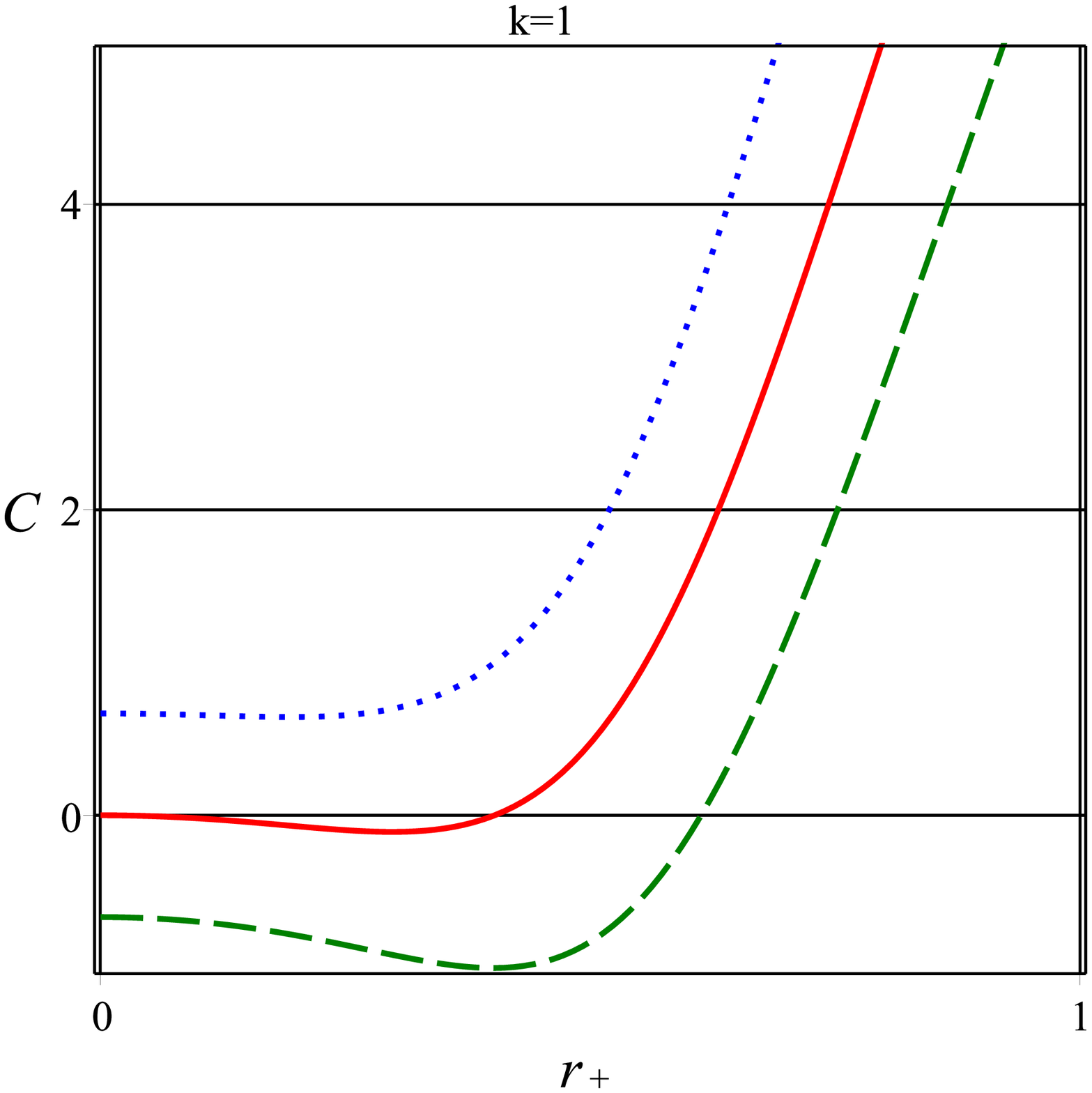}\includegraphics[width=40 mm]{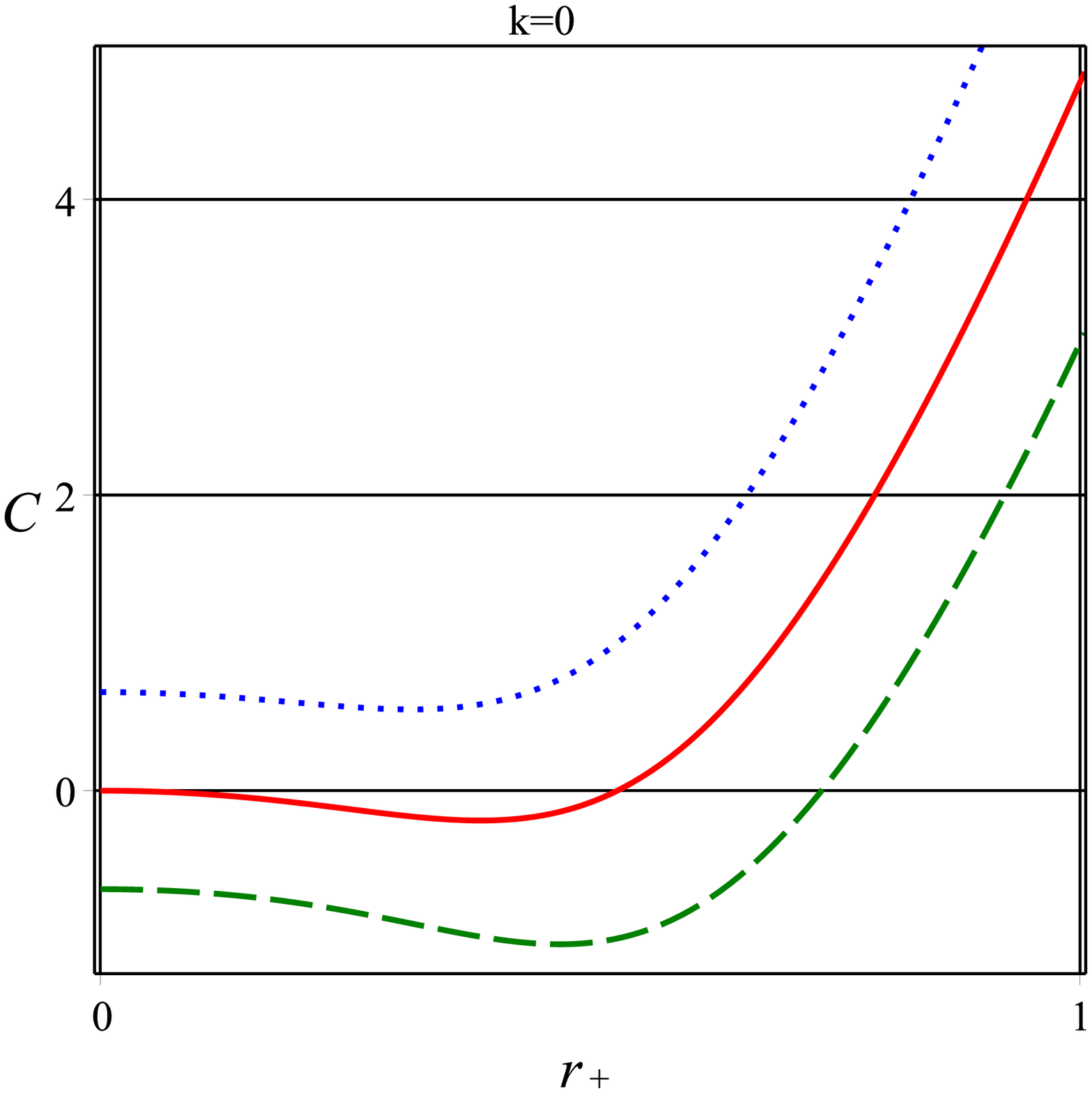}\includegraphics[width=40 mm]{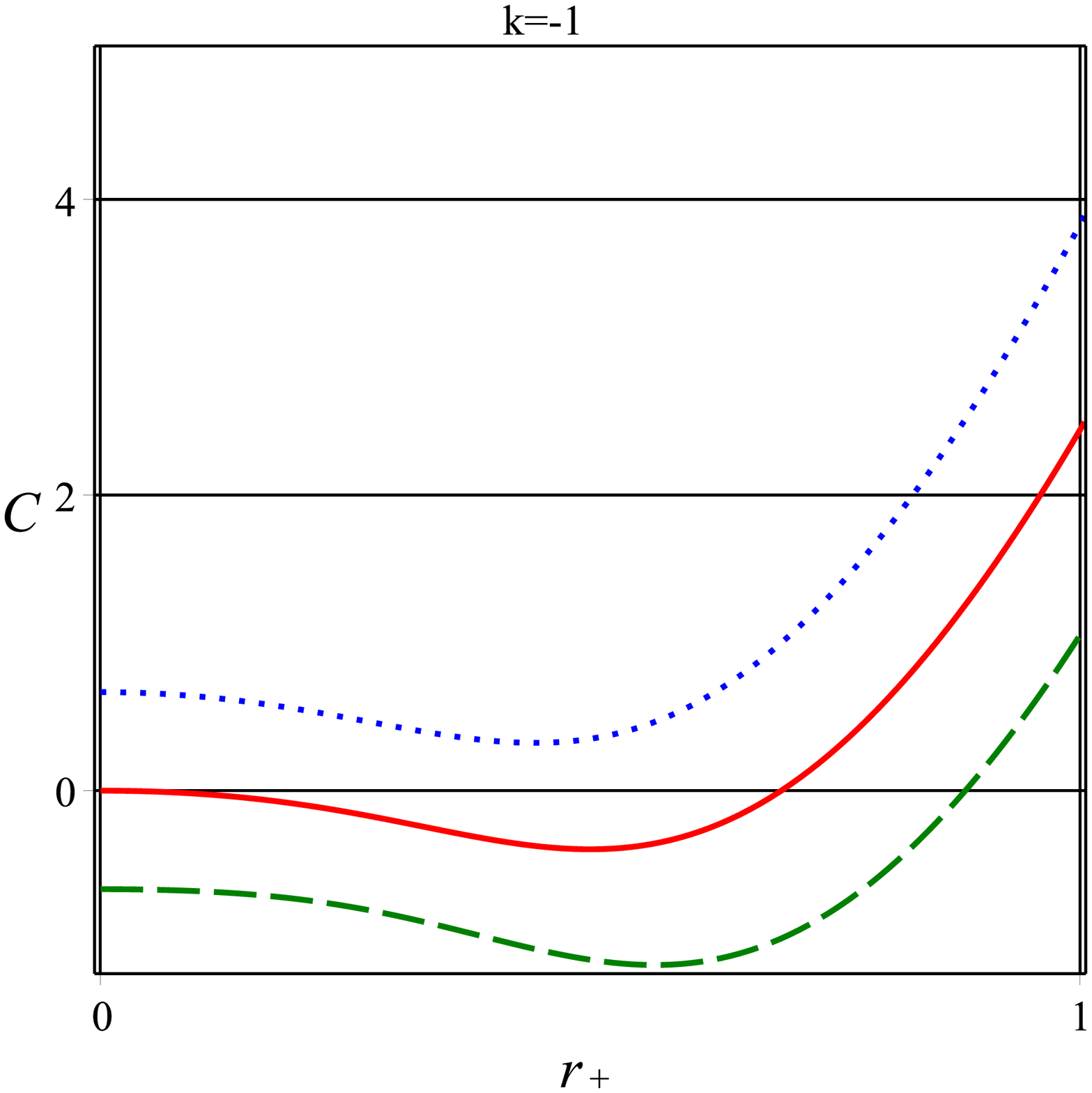}
 \end{array}$
 \end{center}
\caption{Specific heat in terms of $r_{+}$ for $m=0.2$ and all possible values of $k$. We set
unit values for all other parameters.  Green dashed line corresponding to $\alpha=1$, blue dotted line corresponding to $\alpha=-1$, and solid red line corresponding to $\alpha=0$.}
 \label{fig4}
\end{figure}

As illustrated in the previous section, there is no critical point for   AdS black hole of massive gravity in absence of logarithmic correction. Hence, we should consider the effect of   thermal fluctuations to obtain the critical point  and study  the stability of the model. The first step is to study the specific heat which is given by,
\begin{equation}\label{20}
C=T\left(\frac{dS}{dT}\right),
\end{equation}
which  further yields to,
\begin{equation}\label{21}
C=2\pi r_{+}^{2}{\frac {-12r_{+}^{4} -4c_{1}{l}^{2}{m}^{2}r_{+}^{3} -4 
 \left( c_{2}{m}^{2}+k \right) {l}^{2}r_{+}^{2}+{l}^{2}{q}^{2}
}{-12 r_{+}^{4}+4\,{l}^{2} \left( c_{2}{m}^{2}+k \right) r_{+}^{2}-3\,{q}^{2}{l}^{2}}}+C_{\alpha},
\end{equation}
where
\begin{equation}\label{21}
C_{\alpha}={\frac { \left( 4\,c_{1}{l}^{2}{m}^{2}r_{+}^{3}+2\,{q}^{2}{l}^{2}+24\,r_{+}^{4}
 \right) \alpha}{-12\,r_{+}^{4}+4\,{l}^{2} \left( c_{2}{m}^{2}+k \right) r_{+}^{2}-3\,{q}^{2}{l}^{2}}}.
\end{equation}
In the plots of the Fig. \ref{fig4}, we can see the typical behavior of the specific heat at any space curvatures $k=0,\pm1$. We can see that some negative specific heat with $\alpha=0$ and $\alpha>0$, but specific heat is completely positive with negative $\alpha$. It means that the logarithmic correction can remove instability of the black hole. For the positive correction coefficient ($\alpha=1$), the black hole has positive specific heat for $r_{+}>r_{0}$, where $r_{0}$ gives the zero of the specific heat.

\begin{figure}[h!]
 \begin{center}$
 \begin{array}{cccc}
\includegraphics[width=50 mm]{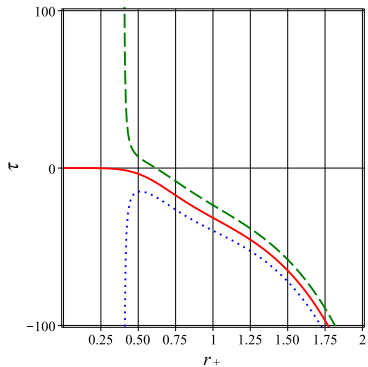}
 \end{array}$
 \end{center}
\caption{Trace of Hessian in terms of $r_{+}$ for $m=0.2$ and we set
unit values for all other parameters.  Green dashed line corresponding to $\alpha=1$, blue dotted line corresponding to $\alpha=-1$, and solid red line corresponding to $\alpha=0$.}
 \label{fig5}
\end{figure}

The important point is that for the charged black holes with chemical potential, the sign of specific heat is not enough to conclude stability of the model and more important test is required using of Hessian matrix of the Helmholtz free energy which we denoted by $\mathcal{H}$, and given by the following matrix,

\begin{eqnarray}\label{mat}
\left(\begin{array}{cc}
\frac{\partial^{2}F}{\partial T^{2}} & \frac{\partial^{2}F}{\partial T\partial \mu}\\
\frac{\partial^{2}F}{\partial \mu\partial T}& \frac{\partial^{2}F}{\partial \mu^{2}}\\
\end{array}\right)=\mathcal{H}.
\end{eqnarray}
By using the relations (\ref{t}),  (\ref{12}) and (\ref{14}), one can find that the 
determinant of the matrix $\mathcal{H}$ vanishes,
\begin{equation}\label{det}
\left(\frac{\partial^{2}F}{\partial T^{2}}\right)\left(\frac{\partial^{2}F}{\partial \mu^{2}}\right)=\left(\frac{\partial^{2}F}{\partial T\partial \mu}\right)\left(\frac{\partial^{2}F}{\partial \mu\partial T}\right).
\end{equation}
It means that one of the eigenvalues is zero and we should consider the other one which is the trace of the matrix (\ref{mat}) given by,
\begin{equation}\label{trace}
\tau\equiv Tr(\mathcal{H})=\left(\frac{\partial^{2}F}{\partial T^{2}}\right)+\left(\frac{\partial^{2}F}{\partial \mu^{2}}\right).
\end{equation}
Now, crucial condition to have stability is $\tau\geq0$. In the Fig. \ref{fig5}, we can see the behavior of $\tau$ with $r_{+}$. It is clear that positive region exists only for the case of positive $\alpha$. Hence, we find that the presence of the logarithmic correction of the form (\ref{s}) with positive $\alpha$ is essential to have critical point and stability at least for small values of the $r_{+}$.\\
For example, we examine the special case with our selected values of parameters. In the case of $c_{1} = 1$, $c_{2} = 1$, $k = 1$, $l = 1$, $m = 0.2$, $q = 1$ we can see that stability exists approximately for $r_{+}\leq0.625$. These values of horizon radius obtained for $m_{0}\leq1.3$. Hence, we find suitable condition on the black hole mass where it is in the stable phase.
 
\section{Conclusions}
In this paper, we have considered a charged black
hole solutions in 4-dimensional massive gravity
with a negative cosmological constant  and studied the first order corrected
thermodynamics and phase structure of the black hole
solutions. In particular, we have computed the first-order corrected  entropy, 
Hawking temperature, Helmholtz function,  internal energy, pressure, enthalpy and Gibbs free energy. We have plotted the
Helmholtz free energy in terms of horizon radius with variation of correction coefficient 
$\alpha$, which confirms that the effect of the logarithmic correction
is important at small $r_+$ (or high temperature) and, also, there exists a critical radius
for which Helmholtz free energy  vanishes.
We found that the logarithmic correction has no important effect on the pressure
of the black hole with large event horizon radius.
However, the  internal energy, enthalpy and Gibbs free energy are the decreasing function of correction parameter. Furthermore, we show that AdS black hole in massive gravity with logarithmic correction is holographic dual of
Van der Waals fluid  for the large black hole and, 
consequently,  found that thermal fluctuations have no important effect. In order to find
effect of thermal fluctuations on the critical points, we have analyzed $P-V$ behavior of
the black hole. Furthermore, we have studied
the effect of thermal fluctuations in order to obtain critical point  and  stability of the model.
From the graphical analysis, we have found that the logarithmic correction
can be used to remove instability of the black hole. However, for the stability of the model, we found remarkably  that trace of Hessian matrix of the Helmholtz free energy must be 
non-negative.

\end{document}